\documentstyle[12pt]{article}

\newcommand{\beq}{\begin{eqnarray}}
\newcommand{\ene}{\end{eqnarray}}

\newcommand{\F}{\noindent}

\newcommand{\MP}{\medskip}
\newcommand{\BP}{\bigskip}

\newcommand{\eq}[1]{(\ref{#1})}

\begin{document}

\rightline{KIMS-1999-11-12}
\rightline{gr-qc/9911060}
\BP

\vskip12pt

\vskip8pt

\Large

\begin{center}
{\bf Quantum mechanical time contradicts the uncertainty principle}
\vskip10pt

\normalsize
Hitoshi Kitada

Department of Mathematical Sciences

University of Tokyo

Komaba, Meguro, Tokyo 153-8914, Japan

e-mail: kitada@kims.ms.u-tokyo.ac.jp

http://kims.ms.u-tokyo.ac.jp/

\vskip8pt

November 17, 1999

\end{center}
\MP

\vskip10pt

\leftskip=24pt
\rightskip=24pt

\small

\F
{\bf Abstract.} The {\it a priori} time in conventional quantum
 mechanics is shown to contradict the uncertainty principle.
 A possible solution is given.

\MP

\leftskip=0pt
\rightskip=0pt

\normalsize

\vskip 8pt

In classical Newtonian mechanics, one can define mean velocity $v$ by
 $v=x/t$ of a particle that starts from the origin at time $t=0$ and
 arrives at position $x$ at time $t$, if we assume that the
 coordinates of space and time are given in an {\it a priori}
 sense. This definition of velocity and hence that of momentum
 do not produce any problems, which assures that in classical regime
 there is no problem in the notion of space-time. Also in classical
 relativistic view, this would be valid insofar as we discuss the
 motion of a particle in the coordinates of the observer's.

Let us consider quantum mechanical case where the space-time
 coordinates are given {\it a priori}. Then the mean velocity of
 a particle that starts from a point around the origin at time $0$
 and arrives at a point around $x$ at time $t$ should be defined
 as $v=x/t$. The longer the time length $t$ is, the more exact
 this value will be, if the errors of the positions
 at time $0$ and $t$ are the same extent, say $\delta>0$, for all $t$.
 This is a definition of the velocity, so this must hold in exact sense
 if the definition works at all. Thus 
\beq
\mbox{we have a precise value of (mean) momentum $p=mv$
 at a large time $t$}
\label{0}
\ene
with $m$ being the mass of the particle. Note that the mean momentum
 approaches the momentum at time $t$ when $t\to\infty$ as the
 interaction of the particle with other particles vanishes
 as $t\to\infty$.

However in quantum mechanics, the uncertainty principle prohibits
 the position and momentum from taking exact values simultaneously.
 For illustration we consider a normalized state $\psi$ such that
 $\Vert \psi\Vert=1$ in one dimensional case. Then the expectation
 values of the position and momentum operators $Q=x$ and
 $P=\frac{\hbar}{i}\frac{d}{d x}$ on the state $\psi$ are given by
$$
q=(Q\psi,\psi),\quad p=(P\psi,\psi)
$$
respectively, and their variances are
$$
\Delta q=\Vert (Q-q)\psi\Vert,\quad
\Delta p=\Vert (P-p)\psi\Vert.
$$
Then their product satisfies the inequality
\beq
\Delta q\cdot\Delta p &=& \Vert(Q-q)\psi\Vert\Vert(P-p)\psi\Vert
\ge|((Q-q)\psi,(P-p)\psi)|\nonumber\\
&=&|(Q\psi,P\psi)-qp|\ge|\mbox{Im}((Q\psi,P\psi)-qp)|\nonumber\\
&=&|\mbox{Im}(Q\psi,P\psi)|
=\left|\frac{1}{2}((PQ-QP)\psi,\psi)\right|\nonumber\\
&=&\left|\frac{1}{2}\frac{\hbar}{i}\right|
=\frac{\hbar}{2}.\nonumber
\ene
Namely
\beq
\Delta q\cdot\Delta p\ge \frac{\hbar}{2}.\label{1}
\ene
This uncertainty principle means that there is a least value
 $\hbar/2(>0)$ for the product of the variances of position and
 momentum so that the independence between position and momentum
 is assured in an absolute sense that there is no way to let
 position and momentum correlate exactly as in classical views. 

Applying \eq{1} to the above case of the particle that starts
 from the origin at time $t=0$ and arrives at $x$ at time $t$,
 we have at time $t$
\beq
\Delta p>\frac{\hbar}{2\delta}\label{2}
\ene
because we have assumed the error $\Delta q$ of the coordinate
 $x$ of the particle at time $t$ is less than $\delta>0$.
 But the argument \eq{0} above tells that $\Delta p\to 0$ when 
 $t\to\infty$, contradicting \eq{2}.

This observation shows that, if given a pair of {\it a priori}
 space and time coordinates, quantum mechanics becomes contradictory.

A possible solution would be to regard the independent
 quantities, space and momentum operators, as the fundamental
 quantities of quantum mechanics. As time $t$ can be introduced
 as a ratio $x/v$ on the basis of the notion of space
 and momentum in this view\footnote[2]{See \cite{K1}, \cite{K2} for 
 a precise definition.}, time is a redundant notion that
 should not be given a role independent of space and momentum.

It might be thought that in this view we lose the relation $v=x/t$
 that is necessary for the notion of time to be valid, if space and
 momentum operators are completely independent as we have seen.
 However there can be found a relation like $x/t=v$ as an approximate
 relation that holds to the extent that the relation does not
 contradict the uncertainty principle (\cite{K1}, \cite{K2}).

The quantum jumps that are assumed as an axiom on observation
 in usual quantum mechanics may arise from the classical nature
 of time that determines the position and momentum in precise
 sense simultaneously. This nature of time may urge one to think
 jumps must occur and consequently one has to observe definite
 eigenstates. In actuality what one is able to observe is scattering
 process, but not the eigenstates as the final states of the process.
 Namely jumps and eigenstates are ghosts arising based on the
 passed classical notion of time. Or in more exact words,
 the usual quantum mechanical theory is an overdetermined
 system that involves too many independent variables: space,
 momentum, and time, and in that framework time is not free from
 the classical image that velocity is defined by
 $v=x/t$.


\vskip12pt
\begin{thebibliography}{8}

\vskip 8pt


\bibitem{K1} H. Kitada,  {\it Theory of local times}, 
Il Nuovo Cimento {\bf 109 B}, N. 3 (1994), 281-302.
 (http://xxx.lanl.gov/abs/astro-ph/9309051,
 http://kims.ms.u-tokyo.ac.jp/time\verb+_+I.tex).

\bibitem{K2} H. Kitada, {\it  Quantum Mechanics and Relativity
--- Their Unification by Local Time}, 
 in ``Spectral and Scattering Theory,"
Edited by A.G.Ramm,
Plenum Publishers, New York, pp. 39-66, 1998.
(http://xxx.lanl.gov/abs/gr-qc/9612043,
 http://kims.ms.u-tokyo.ac.jp/ISAAC.tex, time\verb+_+IV.tex).




\end{thebibliography}
\end{document}